\documentclass[twocolumn]{article}
\usepackage[textsize=tiny]{todonotes}
\usepackage{pgfplots, pgfplotstable}
\pgfplotsset{compat=newest}

\usepackage{amsmath}
\usepackage{natbib}
\setcitestyle{numbers}
\usepackage{hyperref}
\usepackage[margin=0.5in]{geometry}

\AtBeginDocument{%
  }

\begin{document}

\title{Automating GPU Scalability for Complex Scientific Models: Phonon Boltzman Transport Equation}

\author{Eric Heisler$^1$ \and Siddharth Saurav$^2$ \and Aadesh Deshmukh$^1$ \and Sandip Mazumder$^2$ \and Ponnuswamy Sadayappan$^1$ \and Hari Sundar$^1$}

\date{
$^1$University of Utah\\
$^2$The Ohio State University\\[2ex]
\today
}

\maketitle

\begin{abstract}
Heterogeneous computing environments combining CPU and GPU 
resources provide a great boost to large-scale scientific computing 
applications. Code generation utilities that 
partition the work into CPU and GPU tasks while considering data 
movement costs allow researchers to more quickly and easily develop 
high-performance solutions, and make these resources accessible to a 
larger user base.

We present developments for a domain-specific language (DSL) 
and code generation framework for solving partial differential 
equations (PDEs). These enhancements facilitate GPU-accelerated 
solution of the Boltzmann transport equation (BTE) for phonons,
which is the governing equation for simulating thermal transport in 
semiconductor materials at sub-micron scales.
The solution of the BTE involves thousands of coupled PDEs as 
well as complicated boundary conditions and nonlinear processing 
at each time step. 
These developments enable the DSL to generate configurable hybrid 
GPU/CPU code that couples accelerated kernels with user-defined code.
We observed performance improvements of around 18X compared to a 
CPU-only version produced by this same DSL with minimal 
additional programming effort.
\end{abstract}

\section{Introduction}


The analysis of thermal transport is crucial in assessing the performance, cost-effectiveness, and reliability of integrated circuits since overheating is a common cause of their breakdown. In developing methods for heat removal, it is necessary to model the fundamental mechanisms of thermal transport. The size of modern semiconductor devices ranges from a few tens of nanometers to a few hundreds of nanometers. At room temperature, the mean free path of energy-conducting phonons in silicon is approximately 300 nm \cite{Tien1998MicroscaleET}, which is comparable to the device's length scale. In such cases, continuum equations such as Fourier's law of heat conduction are inadequate to accurately predict heat conduction, making it necessary to utilize the Boltzmann Transport Equation (BTE) due to its validity in non-equilibrium heat conduction over a wide range of length scales.

The Boltzmann Transport Equation (BTE) is a seven-dimensional nonlinear integro-differential equation consisting of three spatial coordinates, three wavevector coordinates, and time. Even after linearizing it under the single relaxation time approximation, solving the equation is a difficult task. To address this, researchers have employed the Monte Carlo method, which is useful in incorporating complex physics such as dispersion, polarization, and boundary scattering \cite{mcbte}. However, this method is impractical due to its high cost. Alternatively, deterministic discretization-based methods have been used to obtain the unsteady solution of both gray and non-gray BTE, but the solution process remains a challenging research area in terms of both memory and computational time \cite{mazumder2022boltzmann}.


The BTE poses unique challenges regarding performance, parallelization and scalability  compared with traditional  PDE systems. Firstly, the size of the problem grows rapidly due to its seven-dimensional system, even for modest resolutions. To obtain a spatial and angular grid-independent solution and accurately resolve the length scales, a practical device would require approximately $\sim 10^6$ cells in spatial discretization, $400$ directions ($20$ azimuthal and $20$ polar angles), and $40$ bands in the spectral space ($15$ bands with two polarizations and $25$ bands with one polarization). This typical discretization results in $22000$ coupled PDEs in space and time.
In the time dimension, solvers take $10$-$20$ iterations (depending on the time step size) to attain $3$-$4$ orders of convergence within each time step. In most cases, $\sim 10000$ time steps are necessary to reach a steady state or enable the extraction of time-dependent variables. These computational requirements pose significant challenges in terms of parallelization and scalability, especially strong scalability being a bottleneck for reducing the overall time to solution, as traditional codes can take weeks to months for large 3D problems \cite{mazumder2022boltzmann}. High-throughput architectures, such as modern GPUs can help reduce the overall time to solution greatly speeding up semiconductor design. However,
the complexity of the $7$-D system along with the need to apply complex boundary conditions makes this task extremely challenging.

In this work, we use a modular domain specific language (DSL) to implement the phonon BTE and add support for GPU execution.
The use of a modular DSL enables rapid development of the codes while ensuring correctness and exploration of different work distribution strategies. 
In particular, complexity within BTE applications require specialized boundary conditions that are typically implemented via user-supplied callback functions. These are implemented by domain scientists based on their experimental needs in Julia and are challenging to automatically port to the GPU. 
As a solution to these constraints, we retain such callbacks to execute on the CPU, and optimize for offloading functions to the GPU based on minimizing the overall data-movement between the CPU and the GPU. The DSL automatically partitions tasks between the CPU and GPU by minimizing the data movement. The complexity of the BTE system provides several options for such optimizations. 



\section{Generating Code}
The task of writing optimized code for a complex application is made 
even more challenging when designing it to take advantage of 
heterogeneous architectures. Careful consideration must be made 
about which parts of the computation are best suited to which 
hardware resources. This decision also needs to account for the data
movement costs. Then once an efficient code has been developed any
changes to the model, mesh, or hardware may necessitate a costly
redesign. A good alternative is to generate the code automatically
by combining an abstract description of the problem with details
about the task and computing environment. Now changes to the 
scenario will automatically be accounted for in the generated code.
Here we describe a DSL and code generation framework that is 
capable of generating code for complicated scientific models on
heterogeneous systems.

\subsection{Domain-specific Language}
The DSL Finch is a recently developed tool for numerical solution of PDEs in
the Julia programming language \cite{Heisler23,FinchWebsite}.
It was designed to emphasize flexibility, and includes support for finite 
element and finite volume methods (FEM and FVM). It also has a modular
code generation system that allows development of new generation 
targets aimed at particular software libraries or hardware configurations.

For our application Finch provides key functionality that is lacking
in others. Specifically, we need to efficiently integrate user-written
callback functions into the code, and have the expressiveness to
write a very large number of coupled equations. The ability to easily make
choices about the structure of the generated code, along with the 
ability to hand-modify the code as desired, make this DSL ideal for
exploring and optimizing variations.

In this work we 
focus on an FVM application as described further in the next section. 
More specifically, we consider a large set
of coupled, linear PDEs that will be integrated with an explicit time
stepping method. This means that no linear system needs to be solved,
and the solution is simply advanced in a piece-wise manner. Given the
large number of mostly independent calculations, this problem is very
well suited to a GPU architecture. On the other hand, there are complications
that will require a careful code design as described further below.

Our equation, which is presented in more detail in section 3, is formulated
as a conservation equation for an unknown variable $u$. 
Eq. \ref{eq-conservation} represents a general form of this equation after
integrating over a control volume to make use of FVM. Here $s(u)$ is
a source term. $f(u)$ is a flux that is integrated over the surface of the
cell, and arises after application of the divergence theorem.
\begin{equation}
\int\limits_{V}\frac{\partial u}{\partial t} dV = \int\limits_{V} s(u) dV 
- \int\limits_{\partial V} f(u) dA
\label{eq-conservation}
\end{equation}

Before we can turn this equation into input for Finch, we need
to define the parts of the expression. Variables and coefficients
are represented by entities that have a label, a symbolic representation,
values, and other metedata. Finch uses the SymEngine library to
represent and manipulate the symbolic expressions. It is accessed 
in Julia through the SymEngine.jl library \cite{symenginejl}.
In addition, we need to define some operators to work with the
symbols. SymEngine provides basic arithmetic, while the DSL includes
a set of common differential and vector operators as well as some
special ones like the \texttt{upwind} operator used below. A
powerful feature of the DSL is the ability to define and import
any custom symbolic operator. For example, a more sophisticated
flux reconstruction could be created and used in the input
expression similar to \texttt{upwind}.

The input for equation like (\ref{eq-conservation}) in Finch 
consists of the integrands
on the right side of the equation. Note that the integrals and the time
derivative term on the left are implicitly included.
\begin{verbatim}
conservationForm(u, "s(u) - surface(f(u))")
\end{verbatim}
Here the expressions for $s(u)$ and $f(u)$ need to be replaced with
their actual forms. For example, the reactive source term $-ku$ and 
advective flux term $u\textbf{b}\cdot \textbf{n}$ could be entered as:
\begin{verbatim}
conservationForm(u, "-k*u - surface(upwind(b, u))")
\end{verbatim}
Note that an \texttt{upwind} operator has been defined for reconstructing
an upwind flux of this form.

The first processing step done by Finch is to transform this input
into an expanded symbolic representation:
\begin{verbatim}
-TIMEDERIVATIVE*_u_1 - _k_1*_u_1 - SURFACE *
conditional(_b_1*NORMAL_1 + _b_2*NORMAL_2 > 0,
        (_b_1*NORMAL_1 + _b_2*NORMAL_2) * CELL1_u_1,
        (_b_1*NORMAL_1 + _b_2*NORMAL_2) * CELL2_u_1)
\end{verbatim}

Next, we will consider the forward Euler time integration scheme, though a
similar treatment applies to explicit methods in general. Applying this
scheme results in Eq. \ref{eq-conservation2}, where $u_{0}$ is the
known value from the previous time step.
\begin{equation}
\int\limits_{V} u dV = \int\limits_{V} u_{0} dV + dt \int\limits_{V} s(u_{0}) dV 
- dt \int\limits_{\partial V} f(u_{0}) dA
\label{eq-conservation2}
\end{equation}
The corresponding transformation to the symbolic form results in:
\begin{verbatim}
_u_1 = _u_1 - dt*_k_1*_u_1 - dt*SURFACE *
      conditional(_b_1*NORMAL_1 + _b_2*NORMAL_2 > 0,
        (_b_1*NORMAL_1 + _b_2*NORMAL_2) * CELL1_u_1,
        (_b_1*NORMAL_1 + _b_2*NORMAL_2) * CELL2_u_1)
\end{verbatim}

We approximate these integrals by setting the values in the control
volume or on the surface equal to their average. Then considering a 
polygonal/polyhedral cell with $m$ sides, the calculation for a given cell
is Eq. \ref{eq-conservation-discrete}, where $A_{i}$ is the area of face $i$.
\begin{equation}
u = u_{0} + dt \left( s(u_{0}) 
- \frac{1}{V} \sum\limits_{i=1}^{m} A_{i} f(u_{0}) \right)
\label{eq-conservation-discrete}
\end{equation}
Finch organizes the symbolic terms into categories depending on 
the presence of unknowns and the type of integral. Here LHS refers
to left-hand side, where terms involve unknown variables. RHS
is right-hand side, where all quantities are known. Vol and surf
mean volume and surface integration respectively.
\begin{tabular}{r|l}
LHS vol & \verb|_u_1|\\
RHS vol & \verb|_u_1 - dt*_k_1*_u_1|\\
RHS surf & \verb|-dt * conditional(|\\
 & \verb|_b_1*NORMAL_1 + _b_2*NORMAL_2 > 0,|\\
 & \verb|(_b_1*NORMAL_1 + _b_2*NORMAL_2)*CELL1_u_1,|\\
 & \verb|(_b_1*NORMAL_1 + _b_2*NORMAL_2)*CELL2_u_1)|
\end{tabular}\\

These transformations for time stepping and integration are performed 
automatically by the DSL to match the chosen time stepping scheme
and equation form. Another example is weak form equations
that are used with the finite element discretization. In that
case the terms would be organized into linear and bilinear 
groups, and for volume, boundary, or surface integration.

Once the symbolic representation is expanded, sorted, and 
simplified, it will be combined with the rest of the configuration
information to create a more complete intermediate representation (IR).
Information about the numerical methods and mesh are included to
form a complete description of the computation in the form
of a computational graph. Unlike other such graphs, this IR also
includes metadata about the parts of the computation and comment
nodes to facilitate generation of easily readable code.

The IR must remain at a relatively abstract level to be compatible
with several different code generation targets. Different targets
may perform calculations in different ways and structure the
code in different ways to achieve an optimal solution. This is 
demonstrated in the application in Section 3. Another example is
for linear algebra operations such as matrix multiplication. 
Code generation targets for different languages need to account
for different data layouts to take advantage of vectorization.
This means that the IR needs to represent these computations
on the level of abstract linear algebra operations.

\subsection{Code Generation}
At a high level, the calculation involves a loop over time steps
containing loops over cells to update the values of the unknown 
variable $u$. 
A rough sketch of the needed computation is illustrated here.
Note that \texttt{s(u)} represents the volume integral terms, 
and \texttt{f(u)} the surface integral terms. They will also
include any code for calculating or fetching the needed
coefficients, variable values, and geometric quantities.
For brevity we will only use \texttt{s(u)} and \texttt{f(u)}
\begin{verbatim}
for step = 1:Nsteps
  for cell = 1:Ncells
    source = s(u)
    flux = 0
    for face = 1:Nfaces
      flux += f(u, u_neighbors)
    end
    u_new = u + dt * (source + flux)
    apply_boundary_conditions(u_new)
  end
  u = u_new
  time += dt
end
\end{verbatim}

For our purposes the time step loop is always done sequentially.
The cell loop can be done in parallel very easily, with the only connection
being the need for neighboring values in the flux calculation.
A variety of parallel strategies can be used, ranging from distributed
memory multiprocessing, to CPU multithreading, to GPU multithreading.
Depending on the details of $s(u)$, $f(u)$, and the boundary conditions,
this may be efficiently parallelized on a GPU. When generating code
to run completely on the GPU with one thread per cell, it  has 
the following structure.
\begin{verbatim}
for step = 1:Nsteps
  cell = threadID
  source = s(u)
  flux = 0
  for face = 1:Nfaces
    flux += f(u, u_neighbors)
  end
  u_new = u + dt * (source + flux)
  apply_boundary_conditions(u_new)
  u = u_new
  time += dt
  synchronize_threads()
end
\end{verbatim}

The BTE also involves an additional processing step to evolve 
the temperature in each cell. It is necessary to do this at each
time step because the equilibrium intensity, $I_{o}$, and 
relaxation time, $\tau$, depend on temperature. The relationship
between the non-linear phonon energy distribution and temperature
is highly nonlinear. The change in temperature can be approximated
by calculating the phonon energy flux and using the first law
of thermodynamics. The energy calculation involves
integrating the phonon intensity over all directions and bands, which 
means that all degrees of freedom are loosely coupled for each
cell by this process. 
This relation is not of the form of a PDE that can be expressed 
in the context of Finch. 
This means that hand-written code needs to be developed and coupled
with the code generated by the DSL. 
While Finch supports such user-supplied callback functions, it is
not possible to optimize these the same way as functions written purely in the DSL. 
To complicate things further, this additional
code is written for execution on a CPU, which means that any generated
GPU kernels need to work alongside the host code. Communication is
required, and decisions must be made by the DSL related to the structure
of the code efficiently. Here is one example configuration.

\textbf{GPU kernel:}
\begin{verbatim}
cell = threadID
source = s(u)
flux = 0
for face = 1:Nfaces
  flux += f(u, u_neighbors) [except boundaries]
end
u_new = u + dt * (source + flux)
\end{verbatim}

\textbf{CPU code:}
\begin{verbatim}
for step = 1:Nsteps
  (launch GPU_kernel asynchronously)
  compute_boundary_contribution(u_bdry)
  (synchronize and get u_new from GPU)
  u = u_new + u_bdry
  (external post-processing)
  (send u to GPU)
  time += dt
end
\end{verbatim}

Note that this involves substantial communication between GPU and
host at each time step. In some cases this may be too expensive
to be practical, but as demonstrated below there are cases
in which it is still beneficial. Given the sensitivity of communication,
Finch will automatically determine what variables need to be updated
and communicated during each step. Other values will either only be 
sent once, or not at all.

\section{Demonstration: Boltzmann Transport Equation}
The phonon Boltzmann Transport Equation (BTE) is used to describe heat 
transfer in nanometer-scale semiconductor materials
\cite{Majumdar1997,Majumdar1993}. The challenge of this model is that
it involves solving a seven-dimensional PDE to compute the phonon intensity. 
Typically, this is done by partially discretizing three of the dimensions to
create a large number, hundreds to tens of thousands, of loosely coupled
four-dimensional (time and three spatial coordinates) PDEs as detailed below. 
Although this presents a challenging problem
to solve, it also presents some unusual opportunities for designing efficient 
parallel strategies.

Given the loose nature of their coupling, the equations can be solved
almost independently in parallel. Several parallel configurations have been 
explored with promising results
\cite{Srinivasan2004, Ni2009, Ali2014}. In this paper we will be concerned 
with deterministic methods, though monte carlo techniques have also been
tried\cite{Peraud2011}.

\subsection{Model}
The BTE may be written in terms of the phonon intensity, $I$, as Eq.(\ref{eq-bte}) \cite{Majumdar1997, Majumdar1993}.
\begin{equation}
\frac{\partial I}{\partial t} + v_{g} \cdot \nabla I = \frac{I_{0} - I}{\tau}
\label{eq-bte}
\end{equation}
Where $v_{g}$ is the group velocity, $\tau$ is the scattering time-scale, and 
 $I_{0}$ is the equilibrium distribution function. The intensity represents 
a phonon energy distribution function, 
$I = I(\textbf{x},t,\textbf{s},\omega)$, dependent on position $\textbf{x}$, 
time $t$, wave vector direction $\textbf{s}$, and frequency $\omega$.
The direction and frequency are partitioned into discrete angular vectors
and frequency bands, and will be denoted with the subscripts $d$ and $b$ 
respectively as in $I_{d,b}$.

The frequency space is typically discretized into 40\cite{Saurav2022} to
80\cite{Ali2014} spectral bands. Since we also need to account for longitudinal
and transverse polarizations, and the separate polarizations can also be treated
independently, the number of distinct equations will be larger. For this work we
use 40 frequency bands, which results in 40 longitudinal bands and an
additional 15 transverse bands.
The number of discrete direction vectors for a general 3-dimensional problem
can be around $20\times 20 = 400$\cite{Ali2014}, but simpler configurations 
such as axisymmetric \cite{Saurav2022} and 2-dimensional\cite{Heisler22b} 
can be done with much fewer. For this study, we consider a 2-dimensional 
case with 20 directions.

The finite volume method is used to solve Eq.(\ref{eq-bte}). Integration is
done over a control volume $V$, and the divergence theorem is applied to
the advective term to give Eq.(\ref{eq-bte-fv}). The time derivative can be
integrated using a variety of methods. To facilitate resolution of high-frequency 
transient behavior, the time steps should be kept relatively small\cite{Saurav2022}. 
This means that a simple explicit scheme such as forward Euler is reasonable.
\begin{equation}
\int\limits_{V}\frac{\partial I_{d,b}}{\partial t} dV = \int\limits_{V}\frac{I_{0,b} - I_{d,b}}{\tau_{b}}dV - |v_{g}|_{b} \int\limits_{\partial V} I_{d,b} \textbf{s}_{d} \cdot \textbf{n} dA
\label{eq-bte-fv}
\end{equation}

Temperature is ultimately the quantity of interest, so it is necessary to 
derive thermal information from the phonon intensity. This relationship
is indirect and nonlinear, and must be computed at every time step to
determine the equilibrium intensity. We have adopted the formulation 
and relation to temperature used in \cite{Ali2014} and \cite{mazumder2022boltzmann}. 
Please refer to those for more detailed descriptions of the physics.

The boundary conditions are where the directions may be coupled.
Although intensity of phonons with different directions are treated as
independent in the interior bulk, reflective or symmetry conditions will 
couple values from different directions depending on the geometry of
the boundary. This work includes symmetry and isothermal boundaries,
so the coupling must be considered. Our numerical procedure sets the 
flux through the boundary faces by effectively setting the intensity of 
ghost cells on the outside according to Eq.(\ref{eq-bdry}) where $r$ 
is the direction vector index corresponding to a reflection.
\begin{equation}
I_{d,b}^{outside} = \begin{cases}
 I_{0,b} &\textnormal{at isothermal boundary}\\
 I_{r,b} &\textnormal{at symmetric boundary}
\end{cases}
\label{eq-bdry}
\end{equation}

\begin{figure}[ht]
\begin{center}
\setlength{\unitlength}{0.012500in}
\begin{picture}(115,130)(255,545)
\thicklines
\put(240,545){\framebox(160,130){}}
\put(240,675){\framebox(160,2){}}
\put(240,543){\framebox(160,2){}}
\put(260,665){centered Gaussian heat source}
\put(265,610){initial equilibrium at T=300}
\put(280,550){cold wall at T=300}
\put(195,610){symmetry}
\put(405,610){symmetry}
\end{picture}
\end{center}
\caption{Schematic of the 2-dimensional domain.}
\label{fig-schematic} 
\end{figure}
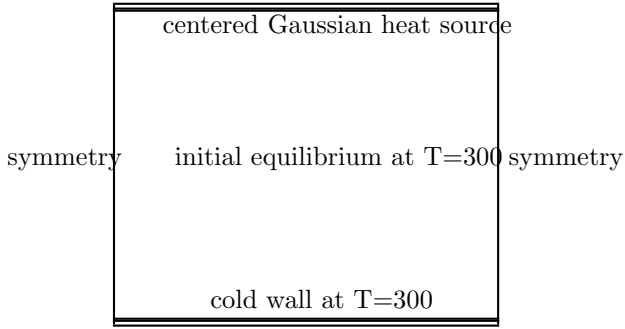

In this demonstration, one side of the domain has an isothermal 
boundary representing a cold wall at the same temperature as the 
initial equilibrium. The opposing wall is also isothermal, but with
a centered heat source with a narrow Gaussian profile. This 
represents a hot spot. 
The remaining boundaries are symmetric, which represents a 
repeating scenario on either side. See figure \ref{fig-schematic}.
The initial condition is a thermal equilibrium at the same 
temperature as the cold wall of the domain. 

\begin{figure}
    \centering
    \includegraphics[width=1.0\linewidth]{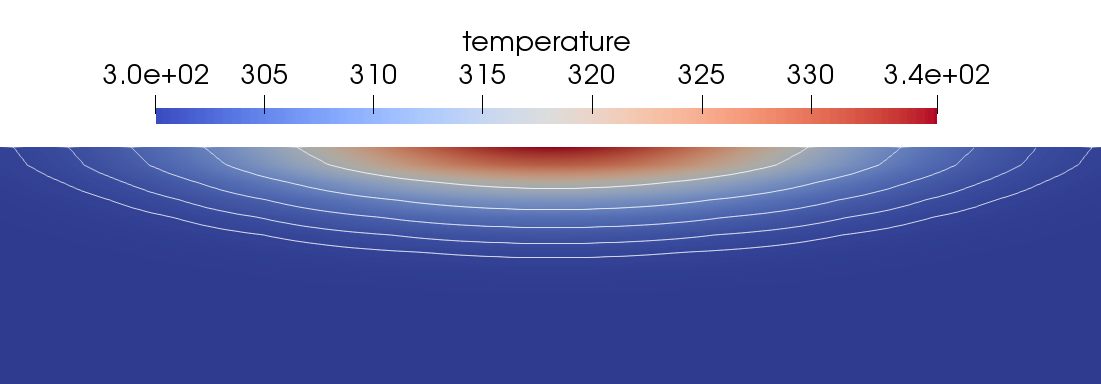}
    \caption{Temperature of the material after 20$\mu s$ (20,000 time steps).
This view is zoomed in to highlight the region around the hot spot.
White contours show the spread of heat.}
    \label{fig-temperature}
\end{figure}

The parameters used in our tests set the initial equilibrium
and cold wall temerature at 300. The hot spot has a peak
temperature of 350 with a $1/e^{2}$ distance of $10\mu m$.
The domain size is $525\mu m \times 525\mu m$ to match
the scenario used in \cite{Saurav2022}. The mesh is a
$120 \times 120$ grid of uniform cells. We use 40 frequency
bands resulting in 55 discrete bands when accounting for 
polarization. A set of 20 uniformly distributed direction vectors
is used, resulting in $20 \times 55 = 1100$ intensity degrees
of freedom per cell, or about $1.6 \times 10^{7}$ overall.
All of the performance calculations below are for 100
time steps, which corresponds to 100ns of elapsed time.
The temperature profile in figure \ref{fig-temperature} 
is after a longer duration of $20\mu s$ (20,000 time steps).

\subsection{Encoding in the DSL}
A goal of the DSL is to take equation input in an intuitive form that
closely resembles the mathematics. The conservative integral form 
of the FVM discretization includes the volume and surface integrals 
of Eq.(\ref{eq-bte-fv}) above. The corresponding input to the DSL
is:
\begin{verbatim}
conservationForm(I, "(Io[b] - I[d,b]) / beta[b] +
    surface(vg[b] * upwind([Sx[d];Sy[d]], I[d,b]))")
\end{verbatim}
Note that an upwind approximation is used for the flux with direction 
vector \texttt{S = [Sx; Sy]}. Since we are using the default flux
reconstruction order of one, this will generate a first-order upwind
approximation.


All of the symbols appearing in this expression, such as variables, 
coefficients, and indices, must first be defined. We refer to these as 
entities, and they are created with the following commands.
\begin{verbatim}
d= index("d",range=[1,ndirs])
b= index("b",range=[1,nbands])
I= variable("I",type=VAR_ARRAY, 
        location=CELL,index=[d,b])
Io= variable("Io",type=VAR_ARRAY, 
        location=CELL,index=[d])
tau= variable("tau",type=VAR_ARRAY, 
        location=CELL,index=[d])
Sx= coefficient("Sx",sx_vals,type=VAR_ARRAY)
Sy= coefficient("Sy",sy_vals,type=VAR_ARRAY)
vg= coefficient("vg",vg_vals,type=VAR_ARRAY)
\end{verbatim}
Although $I_o$ and $\tau$ are assumed as known in the equation 
for $I$, they are dependent on temperature, which is in turn 
dependent on $I$. They are created as variables that have mutable
values for each cell. On the other hand, $\textbf{s}$ and $v_g$ can 
be considered coefficients. They have values that are either 
pre-computed arrays or defined by a function of space-time coordinates.

The initial condition we use for $I$ is the equilibrium intensity for 
a uniform temperature. These frequency dependent values, \verb|I_init|,
are set with
\begin{verbatim}
initial(I, [I_init[b] 
            for d=1:ndirs, b=1:nbands])
\end{verbatim}
The boundary conditions involve a more complicated calculation.
In order to provide more flexibility, Finch allows custom callback 
functions to be imported and used as coefficients and boundary
conditions. This option is a good choice for our problem. Importing
can be done by wrapping the function in a macro:
\begin{verbatim}
@callbackFunction(
  function isothermal(...)
    ...
)
\end{verbatim}
Then the isothermal boundary condition can be set as a flux
condition for variable \texttt{I} on boundary region 1 with:
\begin{verbatim}
boundary(I, 1, FLUX, 
  "isothermal(I,vg,Sx,Sy,b,d,normal,300)")
\end{verbatim}
The relevant values for parameters to be passed to the function
will be interpreted automatically by Finch.

The temperature update that must occur each time step is indirectly
based on the intensity values and involves a nonlinear relation that
cannot be simply written within the context of the PDE for $I$.
Alternatively, we perform it as a kind of post-processing step to be
done after each step. Finch provides a simple way to insert
arbitrary pre-step or post-step code. The following command will
cause the \verb|temperature_update| function to be called
after each time step.
\begin{verbatim}
postStepFunction(temperature_update);
\end{verbatim}

Other configuration details that need to be specified include 
time stepping scheme, discretization type, and order. A mesh 
must either be imported from a Gmsh or MEDIT formatted mesh 
file, or generated internally by Finch's simple generation utility. 
The relevant commands are illustrated in the documentation 
and example scripts are available in the code repository
(reference withheld).

\subsection{CPU Multiprocessing}
The common way of partitioning this sort of calculation is to 
divide the meshed domain into groups of cells to be worked on
in parallel. A significant part of the overhead with this strategy
is the communication of neighboring cell values along the 
partition interfaces. Multithreaded CPU techniques can overcome
this expense, but are limited by hardware constraints on number
of cores and data movement.

\begin{figure}
    \centering
    \includegraphics[width=0.8\linewidth]{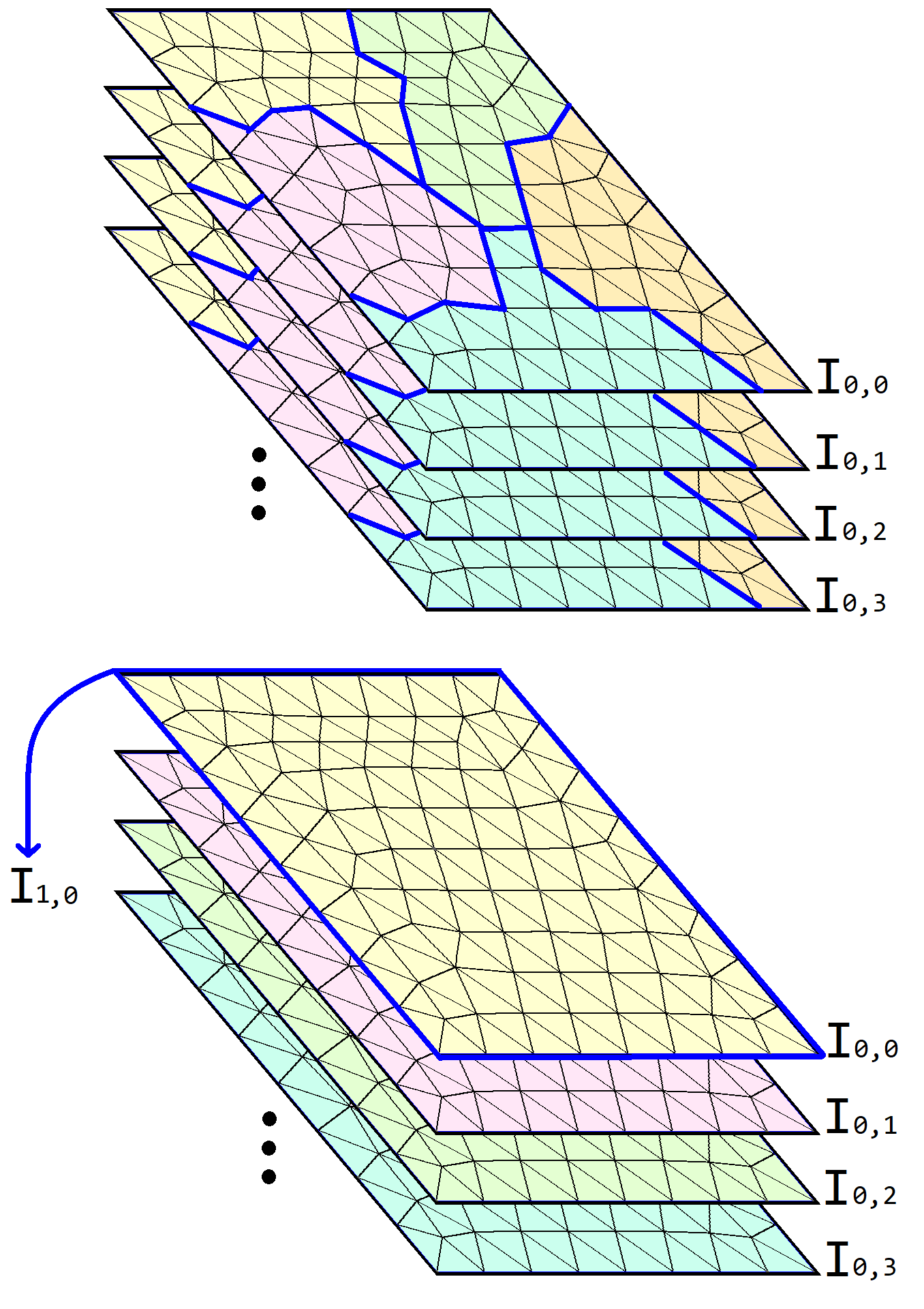}
    \caption{\textit{Top:} Partitioning the mesh requires communication 
		between neighbors for all values of $I_{d,b}$, shown by the blue
  		lines. \textit{Bottom:} Partitioning the equations can require much less
		communication when there are a large number of partitions. 
		In this case, different directions may be coupled
		at the boundary as shown in blue.}
    \label{fig-partitioning}
\end{figure}

The BTE presents another option for partitioning the calculation.
Since there are a large number of equations that are only sparsely
or indirectly coupled, it is more efficient to partition among the 
equations. Figure \ref{fig-partitioning} illustrates the different
communication patterns. Partitioning the equations, or equivalently 
the indices of $I_{d,b}$, requires much less communication. This
is particularly relevant in 3 dimensions and with a large number 
of partitions.

In particular, when partitioning among the bands the boundary 
communication can be avoided as well. The coupling of the bands
only occurs in the temperature update, which in turn only requires a 
reduction of intensity across bands. This would be an optimal configuration
where communication costs are substantial,
but is limited by the relatively small number of bands.

In Finch, the choice of partitioning strategy is simple.
By default the mesh will be partitioned according to the number
of available processes. The library Metis.jl, which is a Julia 
wrapper for the Metis library, is used for mesh partitioning.
Alternatively, one can specify the number of partitions to create
when building or importing the mesh. For the band-parallel case we
will set the mesh partitions to one and instead assign sets of bands
to each process.

When using indexed quantities like $I_{d,b}$, the generated code will
include a set of nested loops like:
\begin{verbatim}
for cell = 1:Ncells
  for dir = 1:Ndirections
    for band = 1:Nbands
      (compute for I[dir,band,cell])
\end{verbatim}
where the default choice of an outermost cell loop is used.
If we wish to permute this ordering to one in which the parallel
band loop is outermost, such as:
\begin{verbatim}
for band = 1:Nbands
  for cell = 1:Ncells
    for dir = 1:Ndirections
      (compute for I[dir,band,cell])
\end{verbatim}
the loop ordering can be set with the command:
\begin{verbatim}
assemblyLoops([band,"cells",direction])
\end{verbatim}
The ability to arrange these loops may also be advantageous
in other applications where efficiency or details of the calculation
favor a particular ordering.

\begin{figure}[!htb]
\centering
\includegraphics[width=1.0\linewidth]{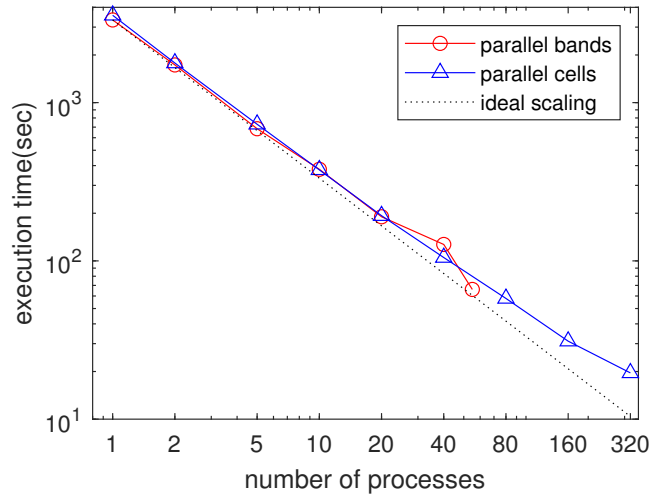}
\caption{Comparison of band-parallel and cell-parallel strategies.
For this configuration the cell-based parallel version is able to
scale to a greater number of processes despite a slightly higher
communication cost.}
\label{fig-cell-vs-band} 
\end{figure}

Performance measurements were done on a cluster
[to be named in final version] with two-socket Intel XeonSP 
Cascadelake nodes with 40 cores each and 192 GB of memory.
Figure \ref{fig-cell-vs-band} illustrates the strong scaling for both
the band-based and cell-based parallel strategies. Despite the
higher communication costs for the cell-parallel version, it was
able to scale well up to 320 processes. 

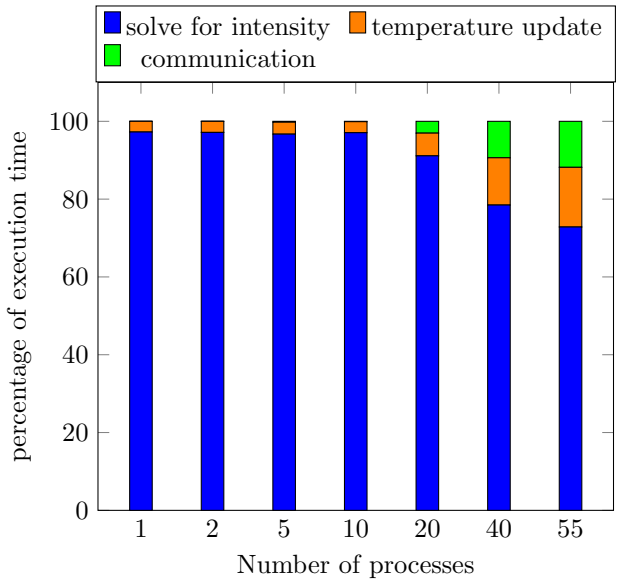
\begin{figure}
\begin{tikzpicture}
  \begin{axis}[
    ybar stacked, ymin=0,  
    bar width=3mm,
    symbolic x coords={1,2,5,10,20,40,55},
    xtick=data,
    xlabel=Number of processes,
    ylabel=percentage of execution time,
    legend columns=2,
    legend style={  at={(0.5,1)},
        /tikz/column 2/.style={
            column sep=5pt},
        anchor=south}, 
    ]
  ]


  \addplot [fill=blue] coordinates {
({1},97.25842754)
({2},97.14934819)
({5},96.70857988)
({10},97.07485407)
({20},91.13192131)
({40},78.48410758)
({55},72.83987024)};
\addlegendentry{solve for intensity}
  \addplot [fill=orange] coordinates {
({1},2.741536712)
({2},2.848848741)
({5},3.069526627)
({10},2.831618667)
({20},5.832442964	)
({40},12.14343928)
({55},15.33470953)};
\addlegendentry{temperature update}
  \addplot [fill=green] coordinates {
({1},0.000035743)
({2},0.001803069)
({5},0.221893491)
({10},0.093527268)
({20},3.035635724)
({40},9.372453138)
({55},11.82542023)};
\addlegendentry{communication}
  
\end{axis}
\end{tikzpicture}
    \caption{Breakdown of execution time for the band-parallel strategy.}
    \label{fig-band-breakdown} 
\end{figure}

A breakdown of the execution time used by different parts of the
calculation is given in figure \ref{fig-band-breakdown}. It is clear
that the calculation of $I$ dominates. For one to ten processes it
accounts for about $97\%$, and even at 55 it takes about $73\%$.
Efforts to improve performance should focus on optimizing this part
of the calculation, which motivates the GPU approach of the 
next section.

\subsection{Accelerating with GPUs}
The parallelization strategy needs to be reconsidered completely
when designing code for a GPU. Communication between threads
becomes less of an issue, and a very high degree of partitioning
needs to be used. However, our band and cell-based parallel 
strategies from the
previous section will still be important when using multiple
GPUs in a distributed configuration.

Rather than generating a set of nested loops, the GPU code 
generator will flatten all of the loops and distribute each degree
of freedom to separate threads. In the interior bulk of the domain
all of the computations have a similar sequence of operations, 
so they can be efficiently computed without thread divergence 
issues. The values on the boundary will require substantially
different work, so should be handled separately. Additionally,
to facilitate more complicated, user-defined boundary conditions,
it is much simpler and more robust to perform this calculation 
on the CPU with the supplied callback functions. 

\begin{figure}[!htb]
\centering
\includegraphics[width=0.9\linewidth]{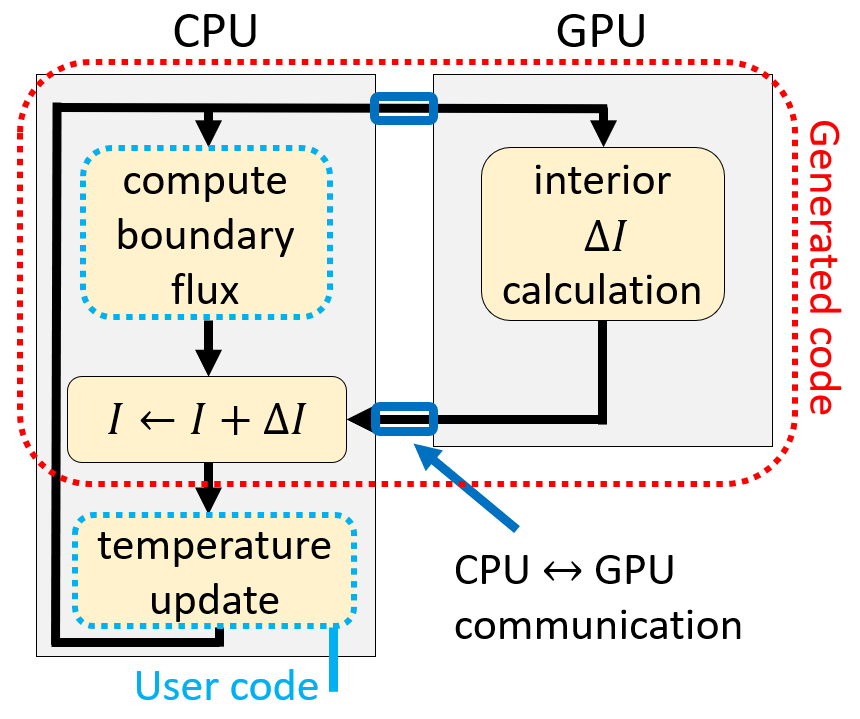}
\caption{Interior and boundary calculations are done asynchronously,
then combined before the temperature update. The variable
values and $\Delta I$ are communicated during each step.}
\label{fig-cpu-gpu-flow} 
\end{figure}

One option is to pre-compute the boundary values and send 
them to the GPU to include in the full calculation. Another
option is to compute boundary contributions 
asynchronously on the CPU to be combined with the interior
part after it has been sent back to the host. 
Since this application requires the values of $I$ to be
sent to the CPU for post-step processing, the communication
will be done either way. Figure \ref{fig-cpu-gpu-flow} illustrates
this procedure.

To accomplish this very different code generation task, one
only needs to instruct Finch to generate code for a GPU
target with the command \verb|useCUDA()|.
This will cause the DSL to look for available GPU resources,
and if available configure the code generation process
accordingly.
Presently this is limited to Nvidia hardware through 
the use of CUDA, with more general GPU support considered
for future development. To interface with the CUDA tools, Finch
uses Julia's CUDA.jl library. This package also provides simplified
ways of allocating and communicating data between the host and GPU.
Profiling macros also allow the interactive use of Nvidia's
profiling software, which was used to determine the performance
metrics given below.

The experiments below were performed with similar CPU configuration
as the previous section paired with eight Nvidia A6000 GPUs per node.
Tests were also done with eight Nvidia A100 GPUs with similar results.
Numerical data used 64-bit floating point numbers. For this application
32-bit numbers did not provide adequate precision for long-duration
simulation.

\begin{figure}[htb]
\centering
\includegraphics[width=1.0\linewidth]{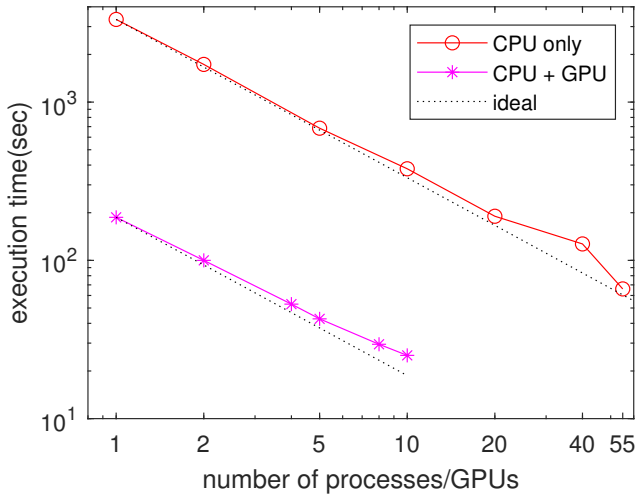}
\caption{Performance of GPU accelerated version compared to 
the CPU-only code. A band-based partitioning is used with multiple
GPUs and the same number of CPU processes. Compared to the
CPU code with an equal number of partitions, the GPU version
is about 18 times faster.}
\label{fig-gpu-scaling} 
\end{figure}

Figure \ref{fig-gpu-scaling} shows execution time for this version 
compared to the CPU-only strategy of the previous section.
The number of GPU devices and CPU processes is set so that each
process is paired with one device. Partitioning between
these is the same as the band-parallel strategy described above.
Strong scaling for this problem configuration is good up to at least
10 devices, but larger numbers did not show further speedup.
Although a direct comparison with the CPU-only code for similar
process counts is not fair, we will state that the best performance
using 20 cores on a single CPU was slightly slower than the same CPU using 
one core and one GPU.
Profiling the computation with one GPU provided the following measurements.

\begin{tabular}{r|l}
SM utilization & 86\% \\
memory throughput & 11\% \\
FLOP performance & 49\% of peak \\
\end{tabular}

\begin{figure}
\begin{tikzpicture}
  \begin{axis}[
    ybar stacked, ymin=0, 
    height=0.38\textwidth,
    bar width=5mm,
    symbolic x coords={1,2,4,8},
    xtick=data,
    xlabel=Number of GPUs/processes,
    ylabel=percentage of execution time,
    legend columns=2,
    legend style={  at={(0.45,1)},
        /tikz/column 2/.style={
            column sep=0pt},
        anchor=south}, 
    ]
  ]


  \addplot [fill=blue] coordinates {
({1},73.40727643)
({2},73.95822823)
({4},71.3876694)
({8},69.74427101)};
\addlegendentry{solve for intensity(GPU)}
  \addplot [fill=orange] coordinates {
({1},25.12993624)
({2},24.4173141)
({4},25.95915251)
({8},25.77216871)};
\addlegendentry{temperature update(CPU)}
  \addplot [fill=green] coordinates {
({1},1.462787333)
({2},1.624457673)
({4},2.653178087)
({8},4.483560279)};
\addlegendentry{communication(CPU$\leftrightarrow$GPU)}
  
\end{axis}
\end{tikzpicture}
    \caption{Breakdown of execution time for the GPU accelerated version.}
    \label{fig-gpu-breakdown} 
\end{figure}
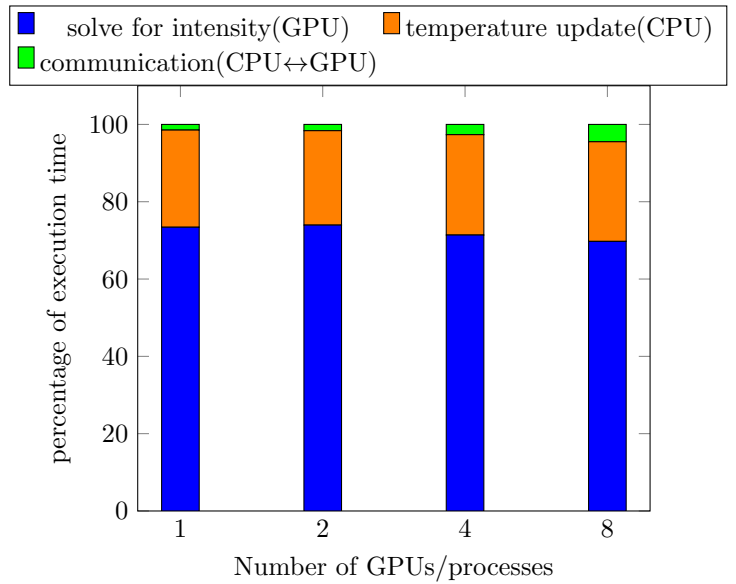

The distribution of execution time is significantly different for this
accelerated version. Figure \ref{fig-gpu-breakdown}, when compared
with figure \ref{fig-band-breakdown}, shows a substantially larger
percentage of time spent on the temperature update. Since that part
of the work is essentially the same for both versions, it is the intensity
calculation that has been sped up. Also note that the communication
time between the GPU and host does not make up a very significant
portion of the time despite the need for communicating variables at 
each time step. Further efforts to minimize communication could have
some benefit, but would not be significant overall.

\subsection{Discussion}

The sections above have illustrated the ease of exploring a
variety of parallel strategies using Finch. When working
exclusively with CPUs, using one of the alternative partitioning
strategies is a good option for reducing communication. The
trade-off is the limited number of possible partitions.
A real benefit of using a band-based parallel strategy comes
into play when working across multiple GPUs, where communication
between devices can be particularly expensive.

The performance benefit of using a GPU to accelerate the calculation
of $I$ is very significant. Combining this with the user-specified
temperature update and boundary condition code, which is run
on the CPU, is simple and seamless from the user's perspective
as the interaction is all handled within the generated code.

\begin{figure}[htb]
\centering
\includegraphics[width=1.0\linewidth]{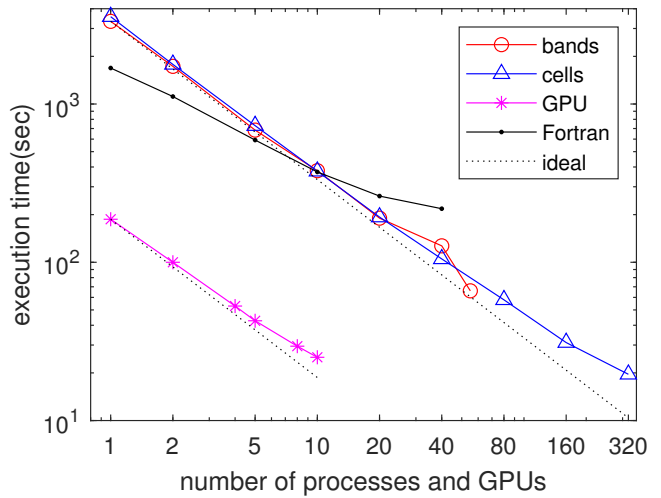}
\caption{Comparison of each strategy as 
well as a reference Fortran implementation based on the same model.}
\label{fig-full-comparison}
\end{figure}

It is also important to compare and verify these results with an 
external code. The exact same model formulation was used by
a previously developed Fortran code that was hand-written and
optimized for band-based parallelism. Our solutions matched
theirs, which had been previously verified against experimental
results. Figure \ref{fig-full-comparison} compares our
performance results against the other code. The sequential
execution of our code takes roughly twice as long as the 
Fortran code. This is reasonable considering the other is a 
hand-written, single-purpose code.

The relatively poor scaling of the Fortran code is due to a slightly
different parallelization of one part of the calculation, which
becomes increasingly significant at higher process counts.
The use of a GPU makes this comparison unfair, but consider
that the GPU version required essentially no additional programming
effort compared to the CPU versions. In terms
of time to solution, we have improved on
the CPU-only codes by a very substantial margin for a given
number of processes. The best possible times were roughly
equal between the 10 GPU run and 320 CPU run. The question
of which version is better depends on the resources available and
the configuration of the problem. When working on a single-CPU
workstation equipped with a GPU, it is clear that the GPU version
would be a good choice.

\section{Conclusion}
The phonon Boltzmann transport equation represents a complicated
PDE that can quickly become computationally challenging due to its
high dimensionality. Along with these complications come unusual
opportunities for parallelizing the problem. We have explored band-based
parallel strategies in comparison to typical cell-based methods. The
development of these codes is facilitated by Finch, which provides
a very simple interface for working with the complex set of equations.

The performance of this computation was improved very substantially
by using GPUs for the expensive part of the calculation. Switching to
this more powerful architecture required essentially no additional
programming effort due to the flexible code generation utilities of
Finch. We were able to improve on the CPU-only codes, including 
a hand-written Fortran code, by a very wide margin.

\begin{figure}
    \centering
    \includegraphics[width=1.0\linewidth]{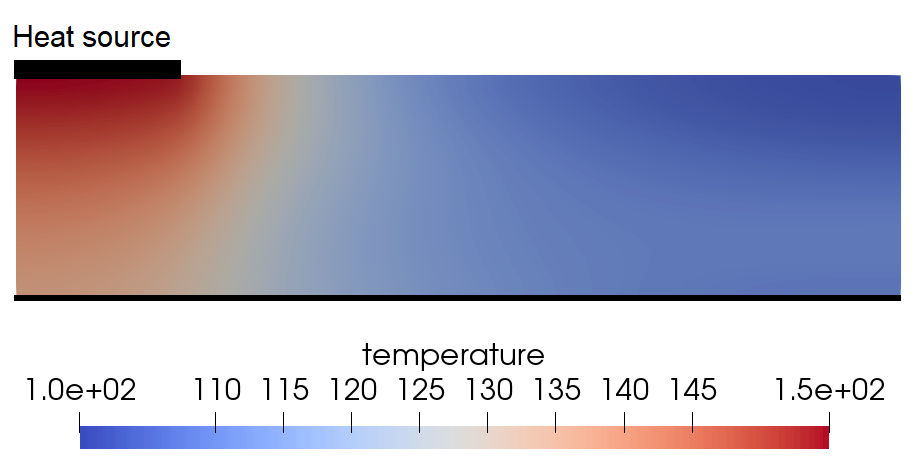}
    \caption{Temperature of a smaller-scale, elongated material with a
    heat source in one corner. Similar to the other example, this has
    symmetry conditions on the left and right, and an isothermal 
    boundary on the bottom.}
    \label{fig-temperature2}
\end{figure}

\section*{Acknowledgement}
This work was supported by grants that are withheld from this review copy to maintain anonymity.

\bibliographystyle{ACM-Reference-Format}
\bibliography{ref-base}

\appendix
\section{Example Input Code}
Below is an example of the Julia input code for Finch. 
This is a simplified version to illustrate the main concepts.
For a full working version, see the example in the repository
\cite{finchRepo}.

\begin{table}[hb]
\textbf{Example input code}
\begin{verbatim}
#=
2D explicit BTE.
=#
using Finch # Load the DSL package
initFinch("bte-gpu");

# Model parameters and callback functions
include("bte-parameters.jl")
include("bte-boundary.jl")

# Configuration
domain(2) # 2-D
solverType(FV)
timeStepper(EULER_EXPLICIT)
dt = 1e-12; nsteps = 10000;
setSteps(dt, nsteps);
useCUDA(); # Tells Finch to generate for GPU

# Import a mesh
mesh("mesh_file.msh")

# Indices and Variables
ndirs = 16;
(t_bands, l_bands) = get_band_distribution(40);
total_bands = t_bands + l_bands;
d = index("d", range=[1,ndirs])
b = index("b, range=[1,total_bands])

I = variable("I", type=VAR_ARRAY,
        location=CELL, index = [d,b])

Io = variable("Io", type=VAR_ARRAY, 
        location=CELL, index = [b])
beta = variable("beta", type=VAR_ARRAY, 
        location=CELL, index = [b])
Sx = coefficient("Sx", dir_x, type=VAR_ARRAY)
Sy = coefficient("Sy", dir_y, type=VAR_ARRAY)
vg = coefficient("vg", group_v, type=VAR_ARRAY)

boundary(I, 1, FLUX, 
    "isothermal_bdry(I,vg,Sx,Sy,b,d,normal,300)")

assemblyLoops(["elements", b, d])

# After each time step the temperature is updated
postStepFunction(post_step()->update_temp(...))

# BTE:
conservationForm(I, 
    "(Io[b] - I[d,b]) * beta[b] + 
    surface(vg[b]*upwind([Sx[d];Sy[d]], I[d,b]))")

solve(I)
\end{verbatim}
\end{table}

\end{document}